# Planning Activities in Software Testing Process: A Literature Review and Suggestions for Future Research

M. Hanefi CALP[1,*], Utku KOSE[2]

[1]*Software Engineering, Of Technology Faculty, Karadeniz Technical University, Trabzon, Turkey*

[2]*Computer Engineering, Faculty of Engineering, Suleyman Demirel University, Isparta, Turkey*

**Abstract**

Software testing process consists of activities that implemented after it is planned and including to document related testing activities. Test processes must be applied necessarily for able to clearly see the quality of software, the degree of reliability, whether it is ready for delivery, the degree of effectiveness and remains of how much testing. One of the most important phase of these processes is test planning activities. Test planning activities directly affects the project's success in software projects. In this study, software testing process, and test planning activities carried out in this process was first clearly demonstrated by the literature review. Later, a basic software testing process and test planning process were determined and explained step by step. In this way, it was aimed to give a different and deep perspective to the test planning activities and to raise awareness on the topic. As a result of the research, it was seen that the test planning activities are not applied adequately at present, software testers, experts or researchers do not have enough knowledge about the details of the method, and this situation causes very serious negative results in the delivery and cost in software projects. Finally, the topic was discussed in detail, and some conclusions and recommendations were given for the personnel in the field of software testing. In this point, the study is expected to contribute the literature.

## 1. INTRODUCTION

At the present time, the inability of software projects to fully meet the requirements has led to the development of large software projects. However, the error rate of such software projects has increased to such an extent that it cannot be ignore. This status demonstrates the importance of software testing activities. In this context, it is necessary to apply the ideal a test process in a fully for the test activities be carried out at maximum quality. As a matter of fact, one of the most important processes of software development activities is testing process. Because, the development of software projects with the lowest error and high accuracy depends on a successful test process. That is, the testing process is a process that directly affects the successful completion of a project [1]. A study by the University of Cambridge in 2013 notes that the global cost is $312 billion a year for detecting and fixing activities of software defects, and it is half of the average project development time [2, 3]. According to various studies, software testing and processes are unfortunately lacking in many companies, and it is usually realize when special circumstances occur. This leads to various negative consequences such as the failure of testing activities in error detection, and time-out in cost and planning [4-6]. Most of the experts, developers or testers agree that an effective software testing process must be implemented to meet the low cost and quality software requirements of software industry. It is also significant for evaluate the performance efficiency of the personnel involved in the test activities in the testing process. This process improves software testing procedures, methods, tools and activities. However, the testing process includes titles such as the continuation of the effectiveness of software product quality, choosing the right test set for the application

*Corresponding author, e-mail: mhcalp@ktu.edu.tr

and making decisions about the optimal sequence of execution of the test and predicting software quality based on the determined defect flaws [7-9].

The testing process of a software project begins with the planning of testing activities. In the planning, issues such as which type of tests will be performed according to the nature of the system to be developed, what actions will be performed for these tests, which test programs will be used, how to report errors to be revealed, how to control software to be tested and how to accept the test, and what trainings they need to carry out these actions, what responsibilities they have within the testing process, where these tests will take place, and how their time will be planned are decided. Therefore, the full implementation of such activities will ensure that higher quality software is obtained. This is important both for the developer and the end user [10].

The aim of this study is to clearly demonstrate the steps, deficiencies of the test planning activities carried out in software testing process and the activities to be need done, and finally to raise awareness on the subject. In addition, a general guideline for experts or testers has been presented, by highlighting little or no applied test planning activities in the real world. In the study, in the second part, the definition and stages of software testing process; in the third part, the importance of test planning activities in the testing process, the steps and tasks followed in this process; then, the discussion section on the subject; finally, in the fifth section, the conclusions and suggestions obtained from the researches are given.

## 2. SOFTWARE TESTING PROCESS

Software testing process consists of a series of actions that were planned, enforced, and the results recorded and documented. This process focuses on the existence of errors in software projects developed [1]. As a matter of fact, one of the most important reasons of software faults is the inadequacy of the test process besides the faulty requirements definitions, inadequate communication between customer and developer, design and code faults, procedural and documentation faults [11, 12]. Jamil et al. discussed software testing life cycle steps and tasks realized during the software testing process (Figure 1). In addition, they said that there is no fixed standard of the testing process in software testing life cycle, and that it shows differences according to regions in the world [13].

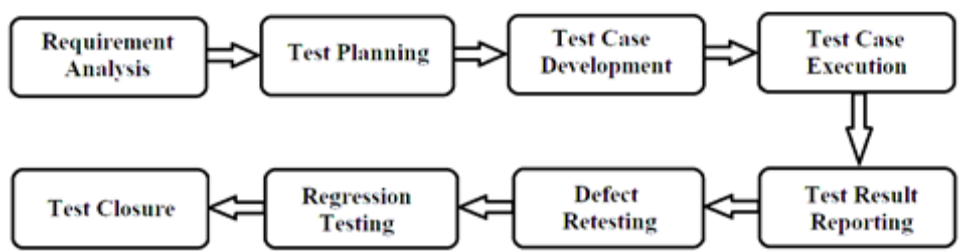

*Figure 1. Software testing process*

In the first phase of the testing process of software projects, software requirements are reviewed, and the basic requirements for testing are performed by software quality team. If any problems arise, the development team takes charge of the task to better understand and solve the problem. Test planning is the second and most important phase of software testing process; this is the step defined by the entire test strategy. In the test designing phase, the test planning activities is stopped. The test cases are usually prepared by the quality team manually or in some cases, automated test cases are formed. Test case defines test data, execution terms, and expected outcomes. The defined set of test data should be chosen such that it generates expected outcomes. This is generally realized to control what terms the application ceases to perform. Test execution phase is consists of execution of the test cases based on the test plan that was produced before the execution phase. All test cases that result unsuccessful are associated with the error or defect. The results of this activity are in the form of a defect report. Finally, test reporting is the document of the created results after the implementation of the test cases, which also consist the defect reporting. Then, this report is forward to the development team and so that defects or errors fixed [13, 14].

According to Garousi et al., testing process is defined as "a test process involves several steps from test planning to test definition (test case designing), execution, and reporting, each of which can be either done manually or automated." [15]. Software testing process is aimed at verification and validation of software. It is a process to keep from activation of error and defects in software or applications. In other words, software testing process is performed to determine defects and errors in software. In this point, software projects should be tested repeatedly until all the defects are found, and they are fixed. Software is controlled and reported on testing process. In Figure 2, there is a software test life cycle suggested by Khan et al. [16].

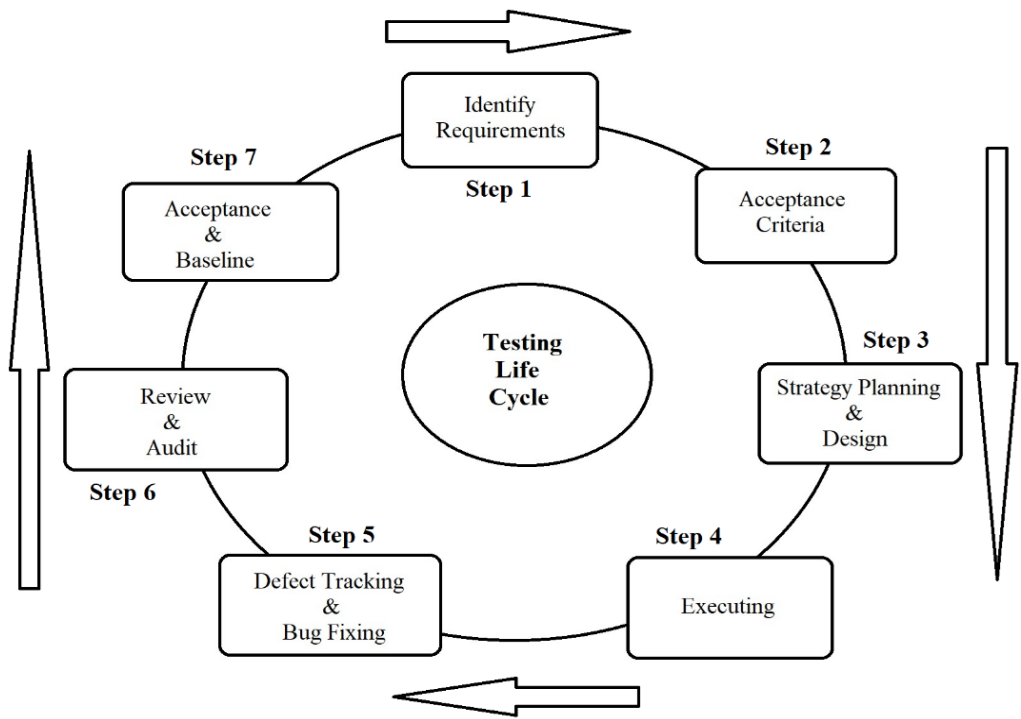

*Figure 2. Testing life cycle*

Software testing process can be realized manual or automated. In this point, the personnel, who is known as a software tester, is a user performing software test to detect unforeseen errors and defects when it is realized manually. In automated test process, the tester uses a test tool and generally codes test code scripts using the JUnit [17, 18]. The testing process in software projects is not just a single activity or process. It includes many sub-activities (requirement analysis, test planning, test case development, forming the test environment, test execution and test closure) and these are briefly called an software testing life cycle (Figure 3) [19].

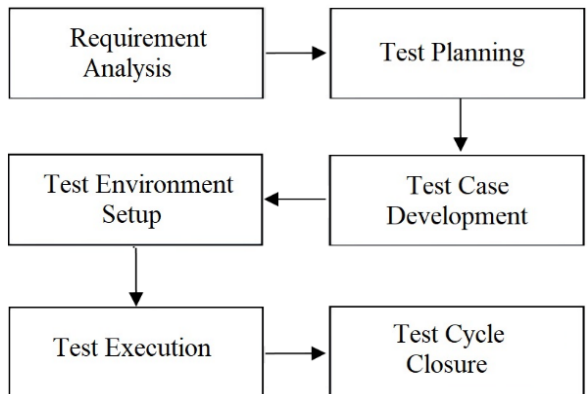

*Figure 3. Software testing life cycle overview*

In this process, it also determines the test tool selection, resource planning and responsibilities, which are distributed among the team members. In addition, it is performed on more technical topics such as architecture, hardware and software test environment. In the test execution phase, the test team will start working according to the test plans, execute the test cases and found software defects are reported. In the test cycle closure phase, which is final phase, software testing staffs come together. They discuss and analyses testing activities in software testing process, and finally detect the most effective methods for future software projects which will developed [19,20].

The maturity of software testing process indicates the quality of the test. A successful testing process includes test plan and risk management, specific test environment and tools, test case definition, test automation, formality on delivery to the test section, test execution, test result analysis, test report, measurement of test effectiveness. In addition, in a successful testing process, the measurements are resumed to determine the effectiveness of the test, and a final status review is performed to determine whether the project requirements have been met after the test has been completed. Test process development is important for the continuous development of the test process, which will increase the effectiveness of test groups. At the same time, test groups uses measurements to determine the current quality of software, the degree of endurance of the current product, the readiness of the product for presentation, the level of quality of the delivered software, the difference between the product quality and the others, the degree of effectiveness of the test applied to software, the problem volume, and how much of the test remains [21-25].

Figure 4 shows the relationship between test cost and errors. In addition, the Figure 4 clearly demonstrates that cost rises in testing both types (functional and nonfunctional). The effective testing is aim to do that optimum level of test activities so that additional testing try can be minimized. According to Figure 4, software testing process is an important component in terms of software quality. The test process is very important. This status can be understand from critical software testing which can be extremely costly because of risks concerning schedule delays, cost overruns, or completely cancel [13, 26-28].

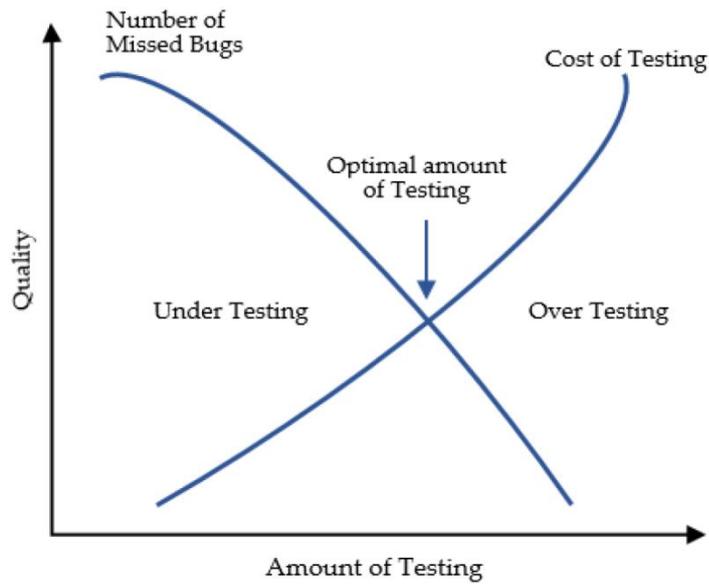

*Figure 4.* *The relationship between test cost and errors*

According to Crespo, generic testing process is consists of phases such as planning, designing of the test case, test execution and test analysis, and test monitoring [29,30]. Most of the researchers tell that there is not a certain testing process, which can be applied to any software project life cycle. Generally, it is observed that each project follows a process parallel to what is shown in Figure 5, even though not all projects track all activities summarized in testing process [31].

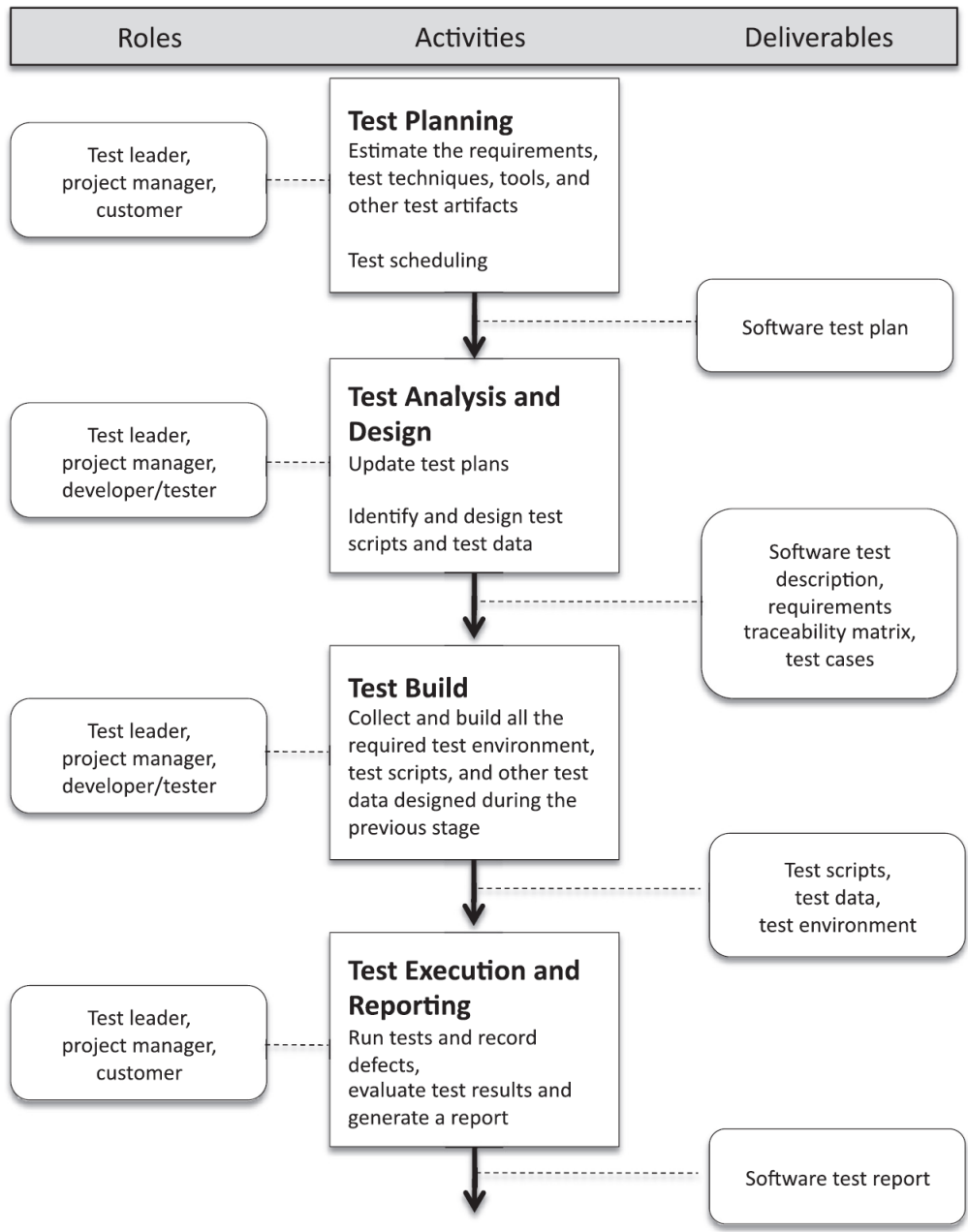

*Figure 5.* Test process

A test strategy of an company define which test process need to be conducted and how they should be performed with development projects with software risks [31, 32]. According to O'regan, software testing process is shown in Figure 6 [21].

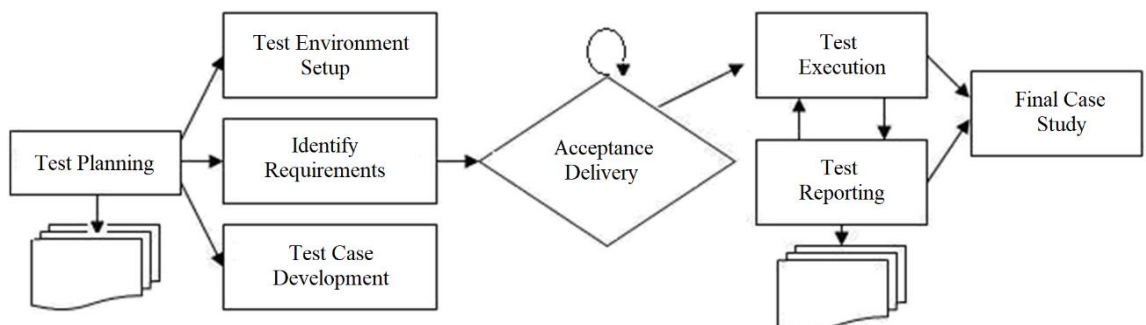

*Figure 6.* Software testing process

As all of the above mentioned processes and activities show, software testing process is an expensive and time consuming process. Integrating this process in the initial stages of the project reduces costs and improves the chances of finding fault. Therefore, it is necessary to first determine the test process successfully and then to treat it with various methods [33, 34].

In Figure 7, the basic test process prepared according to the test processes in the literature is shown and then the activities in this process are briefly explained.

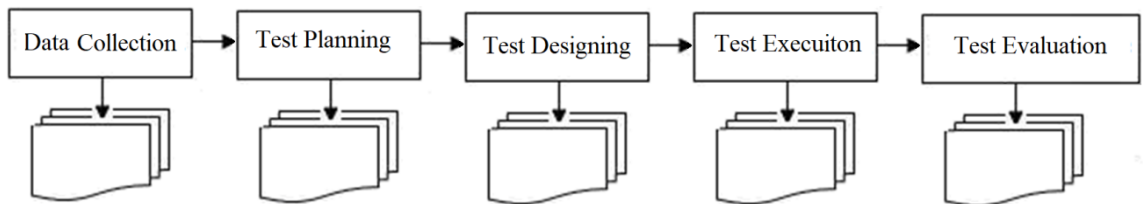

*Figure 7. The simplified test process*

### 2.1. Data Collection

The purpose of data collection is to get information about software development project, to understand the project development field, and to start preparing a test plan. Other information (if necessary) may be collected during project development [35].

### 2.2. Test Planning

A test planning is a document describing approaches such as test activities and services. The test plan is designed to estimate whether software is high quality, understood by everyone, and whether it fulfills its responsibilities. In summary, the aim of the test planning is to establish rules for successful testing in a particular case [21, 23].

### 2.3. Test Designing

Test designing phase begins after the test planning process is successfully completed. Test designing involves primarily the development of specific test requirements and all possible functional varieties. The test designing process is completed by preparing of activities such as the test environment, writing the test cases, and preparing the test procedures [36].

### 2.4. Test Execution

Test execution begins firstly by meeting the test objectives and all the criteria are applied for testing. The tests should be executed according to the test procedures, although with a large number of observed test scenarios and behaviors during the test execution. At the center of test execution activities are found actual results by expected results [37].

### 2.5. Test Evaluation

Firstly, metrics should be analyzed. Following to that, interim reports are published by preparing the reports such as project status report, test error details report, and error report. The test framework plan is created. Finally, the requirements change is determined. In the determining of the first requirement; test functions such as functional composition, functional framework structure, framework standards and system requirements is analyzed [35].

## 3. TEST PLANNING ACTIVITIES IN SOFTWARE TESTING PROCESS

Test planning activities aims to demonstrate what and why will be tested. The input criteria for these activities are prioritized requirements prepared as input for test planning. The supply of this process is the test planning, having predictions and scheduling of resources needed, test outputs to be formed, as well as techniques and methods, tools, and test environments required. The roles are concerned customer, project manager and team leader in this stage. If there is no test team leader in the software project, the coders implements in test planning activities. The exit criterion of test planning activity is the approval of the test plan by the customer and software project manager [31].

Test planning activity is absolutely one of the most important phases in software testing process. This activity involves the subjects of how and what testing will be applied; it allows monitoring, controlling and measuring. The prepared test plan consists of details of the schedule, team members, issues to be tested, and the strategy to be performed (IEEE 2008) [38, 39]. Developers are well warned about what test plans will be executed in the testing process and this information is made usable to project managers and the developers. This test planning is to make them more careful when developing code or making extra changes. In addition, a few organizations have a higher-level document called a test strategy instead of test planning [40].

The first of the actions that directly affecting project success is a good test planning. For this purpose, test plans are prepared at different stages of software projects. In this plans, software components to be tested, the features (functionality, performance, security, usability, etc.), the tasks to be performed, outputs, required resources, responsibilities, timetables and necessary approvals is defined. At the same time, more than one test plan can be produced at different levels to more accurately and effectively identify test actions in projects. For example, the Test Master Plan determining the most general test approach for the Project, or the System Test Plan that defined the system test approach. The following topics are cover when all these plans are developed according to the test plan template specified in the IEEE 829-1998 standards: Test plan descriptor, input, test items, features to be tested, features not to be tested, strategy, successful-unsuccessful criteria's, the criteria's for stopping the test and resuming the test, test outputs, test tasks, environmental requirements, responsibilities, personnel and training needs, calendar, risks and unpredictable, approvals [41, 42].

Test planning can be subdivided into sub-levels that serve a number of purposes. These levels ensure that the planned activities can be easily prepared and understood [43].

In regards to test planning, the following questions are answered; who will be in charge during preparation phase of software testing, establishment of environment, preliminary tests, and testing phase; how to plan the environments where the tests will be carried out; whether a specific application will be made to the project outside of the defined timeline; (if there is a difference) which steps there will be differences from the process; how and when to perform post-test reporting, how to make corrections; how and when to perform post-test reporting, corrections and subsequent testing as needed. The steps to be completed to prepare a successful test plan are;

1. Determine test objectives
2. Develop a test approach
3. Determine the test environment
4. Developing test features
5. Test planning
6. Review and approve the test plan.

An effective testing requires successful planning and execution. The test planning process usually involves the preparation a documented plan by determining the scope and objective of the test to be applied, re-planning in the event of change in objective, determining the test strategy, defining the test environment, identifying any hardware and software resources and tools necessary for the test environment, the effort and research estimate for various actions, identifying risk and probability plans (risk management), monitoring the process and taking right steps, providing status for tests that were passed, blocked and failed regularly, the important events that can be delivered for production, identifying and planning various test types for application, sources, identifying the staff (assigning staff to the tasks) and schedule (time estimate), and case studying to understand any part of software project. Test planning is organized by influential groups to estimate whether software is high-quality, understood by everyone, and it fulfills its responsibilities. Test planning can be revised during the project in a controlled way [21, 45-50]. A simple test plan is shown in Table 1 [21].

*Table 1. A simple test plan*

| Activities | Resource Name(s) | Start Date | Finish/ Re-planning Date | Comments |
|---|---|---|---|---|
| Review the Requirement | Test Team | 01.01.2018 | 16.01.2018 | Completed |
| Comprehensive Test Plan and Review | Test Manager | 13.01.2018 | 28.01.2018 | Completed |
| System Test Plan and Review | Tester 1 | 01.02.2018 | 22.02.2018 | Completed |
| Performance Test Plan and Review | Tester 2 | 22.02.2018 | 31.02.2018 | Completed |
| Usability Test Plan and Review | Tester 2 | 02.02.2018 | 31.02.2018 | Completed |
| Regression Test Plan and Review | Tester 1 | 01.02.2018 | 15.02.2018 | Completed |
| Creation the Test Environment | Tester 1 | 11.02.2018 | 31.02.2018 | Completed |
| System Test and New Test Errors | Tester 1 | 07.03.2018 | 31.04.2018 | Completed |
| Performance Test and New Test Errors | Tester 2 | 18.03.2018 | 07.04.2018 | In process |
| Usability Test | Tester 2 | 18.03.2018 | 15.04.2018 | Completed |
| Regression Test | Tester 2 | 07.04.2018 | 31.07.2018 | Completed |
| Test Report | Test Manager | 01.04.2018 | 31.07.2018 | In process |

The aim of the test planning is to establish a rule for successful testing in a given event. In this point, the most important issue is the documentation. Because documents make it possible to manage software projects better. Test execution and analysis are more easily implemented when a test plan is prepared in detail and fully. The test plan is an evolving document, especially because the system is continuously changing in the spiral environment.

A successful test plan:

- are high the chances of detecting the majority of faults,
- provide test coverage for many of the codes,
- is flexible,
- is easily executed and repeated,
- determines the test types to be applied,

- clearly documenting the expected results,
- provides an opportunity to correct the error in the event of an error,
- clearly define the test objectives,
- uncovers the test strategies,
- clearly define test input and output criteria,
- not much more than necessary,
- diagnoses the risks,
- specify test requirements,
- specify the deliverability of the test [46, 51].

Therefore, the test planning activity within the test process is a very important step in order to manage the requirements accurately and completely. The main and sub-activities in the test planning process can be explained as follow:

### 3.1. The Main and Sub-Activities in Test Planning Process

The main and sub-activities in Table 2 must be performed to ensure that the test planning process is successful.

*Table 2.* *The activities of test planning process [35]*

| Main Activities | Sub-Activities |
| --- | --- |
| Creating a Test Plan | Preparing an introduction |
| | Determining high-level functional requirements |
| | Determining manual / automatic test types |
| | Identifying the test exit criteria |
| | Preparing regression test strategies |
| | Determining the test deliverables |
| | Organizing the test team |
| | Preparing the test environment |
| | Determining the dependencies |
| | Preparing a test schedule |
| | Choosing the test tools |
| | Preparing defect recording / monitoring procedures |
| | Preparing change request procedures |
| | Preparing version control procedures |
| | Determining the configuration build procedures |
| | Identifying project problem resolution procedures |
| | Preparing reporting procedures |
| | Determining approval procedures |
| Determining the Metric Objectives | Determining the metrics |
| | Determining the metric points |
| Review / Approve the Plan | Reviewing/ approving the plan |
| | Obtaining approvals |

### 3.1.1. Creating a test plan

There are two methods to prepare a test plan. The first approach is a main test plan that shows an each test plan in detail. A detailed test plan validates certain stages in the waterfall development life cycle. Examples of test plan consist of the unit, integration, system and acceptance tests. Other test plans prepared in detail involve application upgrades, regression testing and package installations. Unit test plans are code-centric and very detailed. However, it is short because it has limited fields. The system or acceptance test plans focus on the black box look or functional test of the entire system, not software unit. The second approach is a test plan. This approach is often referred to as an acceptance/system test plan, but is a test plan including all of the planning considerations for unit testing, integration testing, system testing and acceptance testing [21, 42].

In order to complete that, main and sub-activities that must be fulfilled are as follows:

*Preparing an introduction*

In this section, risks of the application, possibilities, objectives / targets, benefits, and important success factors of the company are documented. Introduction preparations include goals such as recent product offerings, improved productivity (internal and external), growing, organizational vision, financial (income, costing), competitive situation and market leadership.

*Determining high-level functional requirements*

Documenting the functional requirements of the system as a text or use-case model is a step in the process. High-level functional requirements must be explicitly defined before the system can be passing to the decomposition phase [52]. The fundamental functional list includes the master functions of the system and is defined by the verb-object paradigm [35].

*Determining manual / automatic test types*

Test types such as usability, security, performance and regression are determined according to the objectives or targets of the application. For example; if the application has a financial structure used by most of people, certain security and usability testing must be enforced [36, 53].

*Identifying the test exit criteria*

The test exit criterion is a part of the preparation process for testing activities within the test process. These criteria indicate the completion of the test phase and are the conditions that must occur before testing phase. Termination of the test without exit criteria can be defined as the depletion of the time or resources required continuation. This tells about the quality of the system and can be a sign of the test quality of the application [54]. Ending of test time scheduled for test exit, finding a lot of predefined defects, all official tests must be run without any errors being identified, and all of the above mentioned must have occurred [35].

*Preparing regression test strategies*

Regression testing is a type of software test that investigates whether to introduce new defects or setbacks after changes that make up a system, such as functional developments, patches, or configuration changes [55]. The regression test tests the application's evolution cycle, defect correction and removal of changes that occur during maintenance. The test that cannot find an original defect or error should be run again after the defect was fixed. Much effort should be devoted to situations where the original defect or error is corrected and the symptoms are incomplete. The other test cases in the functional field where the defect is not discovered must be in the regression test chapter. Customer reporting defects should be prioritized and should be subjected to regression testing from beginning to end.

*Determining the test deliverables*

Test deliveries are the results of test planning, test designing, test development, requirement change, test case, metrics, total test record report, test case record, interim test report, total system report, defect report and test defect documentation. That is, the reports and documents indicate the deliverability of the developed software. The test manager identifies the delivery status based on these documents.

*Organizing the test team*

The test team should provide staff at the highest level. Test team managers and test team should be motivated to work on the project, and create opportunities to demonstrate their experience [35].

*Preparing the test environment*

The test environment is software and hardware that the test team builds to test the newly developed software product. The aim of the test environment is to provide the physical environment required for test activities. Accordingly, the needs of the testing environment are determined and re-gauged before implementation [56, 57].

*Determining the dependencies*

It should be identified test dependencies such as suitability of the code and tester, test requirements, suitability of the test tool, test group training, technical support, timely defect detection, adequate test duration, computers and related documentation, documentation related to system, development method, accessibility of the test lab field and development and contracts (procedures and processes).

*Preparing a test schedule*

Test schedule should be created to included test steps, targeted beginning and ending dates and responsibilities. This schedule should be define how the interview, monitoring and approval it should be [35].

*Choosing the test tools*

Software testing tools play an important role in creating a more accurate and fast software product by providing an automation approach especially on large projects, using time and work efficiently and increasing quality. Software tools are important for effective testing. The selection of test tools is based on feeling or judgment. But, more systematic approaches can be also performed [1, 58,59].

*Preparing defect recording / monitoring procedures*

Defects are detected during the test process, these defects must be recorded and a defect report form must be designed. In most of the defect form, there are the identifying of the problem (functional field, type of problem, etc.), the quality of the problem, the conditions causing the problem (inputs and steps), the environment in which the problem occurs (platform), the diagnostic information (defect code etc.) and effect of the problem (results) [35]. These activities need to be controlled using defect tracking tools. Because, these tools are effective in tracking of software, recording of the process, finding and managing of the defects in developed software. There are many paid or free defect-tracking tools in the literature, including desktop applications and web-based applications [60].

*Preparing change request procedures*

If everything was perfect, a system would have been established and there would have been no change. However, nothing is perfect, and situations such as he changing in requirements after system is developed, change requests, design changes, incomplete and unclear features, errors that cannot be discovered during the examination, changes in software environment (platform, hardware, etc.) can be encountered. These situations require new regulations to be made [25, 61].

*Preparing version control procedures*

The only method of defining each software component is by project tagging. Each software component must have a single name. Delivery and level numbers such as 1.1, 1.2 are given in each arrangement of software components. Software components are numbered 1 when they are first defined, and 2, 3 on subsequent ones respectively.

*Determining the configuration build procedures*

The configuration procedures need to be detected in order to determine the component models and to run the component building techniques. The configuration structure model reveals important questions such as how to control the road elements. The configuration typically provides a set of descendants from software components.

*Identifying project problem resolution procedures*

Problem management procedures must be set up before the project starts. Procedures should address how a problem is reported, showed (rejected, deferred, merged or accepted), investigated, approved, delayed, rejected or abolished.

*Preparing reporting procedures*

Test reporting procedures are important for managing the testing process and the expectancies of software project team members. The purpose of the test case reporting process is to reporting the testing process in the direction of the objectives, the testing issue, problems, and interests of the report. There are two important reports that need to be prepared.

1. Interim test report: This report shows the status of the test effort.
2. Total system report: The total report of a test covers the test report after the completion of the whole test.

*Determining approval procedures*

Approval procedures are important in the test project. They help to ensure the necessary agreement between project team members. The approval procedure can transform with comments from an official test to an informal review [35].

### 3.1.2. Determining the metric objectives

Completing of this process is possible with two tasks: "Determining the metrics" and "Determining the metric points".

*Determining the metrics*

Several metrics need to be collected in order to be able to tracking, improve the processes applied during software life cycle and to determine that the product is at the desired quality level [62]. In other words, the collection of product and process metrics is an activity that must be carried out in order to measure both the quality of the product and the effectiveness of the processes [63].

*Determining the metric points*

Determination of metric points plays an important role in achieving metric targets [62].

### 3.1.3. Review / approve the plan

This process have two tasks. These: "Reviewing/ approving the plan" and "Obtaining approvals".

*Reviewing/ approving the plan*

The test plan reviewing should be well planned in the progress of the actual review and the participants should take the final copy of the test plan. The purpose of this task is to develop, to agree on the project sponsor, and to accept the test plan [35, 63].

*Obtaining approvals*

Approval is important in helping to provide necessary agreements between testing, development and sponsorship in the test effort [35].

## 4. DISCUSSION

This section contains important points and findings from the study process. In this point, this study is clearly reveal that there are many steps to be taken in the test planning process, each of these steps requires a resource in terms of both cost and time and it is uncertain what level of success will be achieved. This status is a serious problem for test managers. However, test planning process in software projects is very expensive, laborious, difficult and complex. Therefore, if the test planning activities are not fully performed, the testing process is failing.

The test planning phase is usually considered the first phase of software testing process. However, when looking at software testing processes in the literature, in almost all processes, the requirements are first defined or required information is gathered, then test planning is resumed. This does not indicate that the test planning phase is unimportant. On the contrary, test planning activities are very important because they start in the early stages of the testing process, and should be especially prepared very well.

Test planning is one of the most critical phase of software development lifecycle because of it is dependent the recent delivery of the product. In this point, test planning process is take very time, thus developed techniques, advanced and original methods are necessary in this process. It can increase the available test planning methods, both for time effectiveness as well as for effective and dependable recent product which not only materials the specified requirements but also ensures with maximum efficiency.

The test planning phase also contains applications that address the definition of the test environment. In this content, the test environment requirements should be reveal, and analyze by ranked as compulsory and inter-related.

In addition, test planning activities can be performed automatically in order to save time and cost. However, test automation is not also a final decision; thus, it should be appropriately decide for obtain the successful results (optimized cost/benefit of testing) in the application of test activities. Today's, although software test planning activities is a highly investigated subject in the field of software engineering, there is not much method to detect what this activities are worthwhile automating, and what activities are not. The experimental researches or experiments are required to find the optimum composition of manually realized and automated test planning activities, and these investigations should be further deepened. Additionally, making estimates of the results of software test planning activities using artificial intelligence methods and analyzing these results will positively affect the success of these activities.

## 5. CONCLUSIONS AND RECOMMENDATIONS

In this study, firstly, the definition and phases of software testing process are explained, and the test planning process from software testing process stages has been extensively investigated. At the same time, information that will eliminate the misconceptions and incorrect applications about the subject were included. In the investigated studies, it had been obtained that the test planning activities are not applied adequately at present, experts do not have enough knowledge about the details of the method, this situation causes very serious negative results in the delivery and cost in software projects, and especially the testing of software projects is very important for researchers. It was also found that, the test planning methods applied in very small numbers, did not occur in the required sequence or that the personnel involved were not serious about the planning activities.

In addition, there are also some suggestions that should be taken into account during the test and test planning process, which is important for software test engineers in the study. These suggestions can be summarized as follows: Firstly, an ideal test process should not be considered separately from the coding process. Then, the test plans specifically highlighted should be prepared in the first stages of the development process and should develop the interactions of analysis, design and coding activities. The test plan should clearly reveal the test goals, area, strategy and approach, testing procedures, testing environment, test completion criteria, test cases, components to be tested, tests to be applied, test timetable, personnel requirements, reporting procedures, estimates, risks and probability planning. It should be ensured that the test plan is accessible to the appropriate individual and available when a test plan is being developed. Finally, the steps and tasks that are widely discussed in the study should be implemented as much as possible in order completion successfully of the test planning process, or at least for optimum success.

**CONFLICT OF INTEREST**

No conflict of interest was declared by the authors.